\documentclass
[aps,prl,a4paper,superscriptaddress,twocolumn,showpacs,lengthcheck,nobibnotes,nofootinbib,floats,balancelastpage,10pt]{revtex4}%
\usepackage{amsfonts}
\usepackage{amsmath}
\usepackage{amssymb}
\usepackage{graphicx}%
\begin{document}
\title{Ion-beam driven dust ion-acoustic solitary waves in dusty plasmas}
\author{N. C. Adhikary}
\email{nirab_iasst@yahoo.co.in.}
\affiliation{Material Sciences Division, Institute of Advanced Study in Science and
Technology, Paschim Boragaon, Garchuk -- 781 035, Guwahati, Assam, India.}
\author{A. P. Misra}
\email{ap.misra@physics.umu.se.}
\affiliation{Department of Physicss, Ume\aa \ University, SE-901 87 Ume\aa , Sweden. }
\author{H. Bailung}
\affiliation{Material Sciences Division, Institute of Advanced Study in Science and
Technology, Paschim Boragaon, Garchuk -- 781 035, Guwahati, Assam, India.}
\author{J. Chutia}
\affiliation{Material Sciences Division, Institute of Advanced Study in Science and
Technology, Paschim Boragaon, Garchuk -- 781 035, Guwahati, Assam, India.}

\begin{abstract}
The nonlinear propagation of small but finite amplitude dust ion-acoustic
waves (DIAWs) in an ion-beam driven plasma with Boltzmannian
electrons,  positive ions and  stationary charged dust grains, is
studied by using the standard reductive perturbation technique (RPT). It is
shown that there exist two critical values $(\gamma_{c1}$ and $\gamma_{c2})$
of ion-beam to ion phase speed ratio $(\gamma)$, beyond which the
beam generated solitons are not possible. The effects of the parameters,
namely $\gamma$, the ratio of the ion-beam to plasma ion density $(\mu_{i})$,
the dust to ion density ratio $(\mu_{d})$ as well as the ion-beam to plasma
ion mass ratio $(\mu)$ on both the amplitude and width of the stationary DIAWs
are analyzed numerically, and applications of the results  to laboratory ion-beam as well as space plasmas (e.g.,
auroral plasmas) are explained.
\end{abstract}
\date{26 Oct., 2010}
\startpage{1}
\endpage{102}
\maketitle


Low-temperature plasmas containing massive charged dust particles are
frequently found in various space plasma environments \cite{Whipple} as well as in
laboratory devices \cite{Sheehan} and industrial processes \cite{Selwyn} in
the form of complex plasmas. The presence of these highly massive and
negatively charged dust particles in an electron-ion plasma is responsible for
the appearance of new types of electrostatic waves including solitary waves,
depending on whether the dust grains are considered to be static or mobile.
These electrostatic solitary waves have already been observed throughout the
Earth's magnetosphere at the narrow boundaries, e.g., the plasma sheet
boundary layer and the polar cap boundary layer \cite{Matsumoto}. One of these
solitary waves is the dust ion-acoustic (DIA) wave, which is the usual
ion-acoustic wave (IAW) modified by the presence of static dust grains. During
the past several years, after the theoretical prediction of the existence of
such DIA waves (DIAWs) by Shukla and Silin \cite{Shukla}, and their
experimental verification by Barkan et al \cite{Barkan}, extensive works have
been devoted to study the features of such DIAWs by many authors both
theoretically (see e.g., \cite{Bharuthram}) and experimentally (see e.g.,
\cite{Nakamura1}).

On the other hand, it has been found that sufficiently energetic charged
particles like ion-beams can significantly affect the propagation
characteristics of solitary waves in plasmas \cite{Okutsu}. In the auroral
zone of the upper atmosphere, such types of solitary structures have been
found in the vicinity of ion-beam regions usually having negative potentials
\cite{Temerin}. The spacecraft observations in the Earth's plasma sheet
boundary layer show the existence of both electrons and ions in the range of
keV energy. Observations also indicate that both of these ions and electron
beams can drive the broadband electrostatic waves present there \cite{Parks}.
However, these ion beams in laboratory dusty plasmas have become indispensable
in the field of materials processing such as etching chemical vapour
deposition and surface modification \cite{Sugai}.

A very few theoretical works on the behaviors of solitary waves in
multi-component ion-beam plasmas have been done by some researchers ( see,
e.g., \cite{Abrol}). It has been estimated that the presence of ion beams
plays an important role in breaking up the solitary waves into many more
solitons \cite{Abrol}. The properties of solitary waves under the influence of
high speed as well as slow ion beams on the propagation of IAWs have also been
investigated both theoretically and experimentally \cite{Nakamura2}. However,
the study of DIAWs under the influence of ion-beams, which may often exist in
the space plasma environments (e.g., in the auroral regions), has not yet been
reported in detail. Moreover, though the role of charged dust grains in
the auroral region has not yet been directly established, they could be
important from theoretical view-points of space plasmas. Thus, the study of
DIAWs under the influence of charged dust grains as well as ion-beams could be
of interest to observe the ion wave oscillations in laboratory as well as
space plasmas. This is the basic purpose of the present brief communication.

We consider an unmagnetized ion-beam driven dusty plasma composed of positive
plasma ions, positive ion-beams, Boltzmann distributed electrons and
negatively charged dust grains forming only the background plasma. Two
distributions for ions: one is the bulk, uniform cold ion plasma with its
equilibrium flow speed equal to zero and other the energetic ion component,
i.e., the ion beams  having equilibrium ion-beam speed $v_{b}^{(0)}$, have
been considered for the present system. The normalized set of basic equations
describing the propagation of DIAWs is $\partial_{t}n_{\alpha}+\partial
_{x}\left(  n_{\alpha}v_{\alpha}\right)  =0,$ $\partial_{t}v_{\alpha
}+v_{\alpha}\partial_{x}v_{\alpha}=-\zeta_{\alpha}\partial_{x}\phi$ and
$\partial_{x}^{2}\phi=\left(  1+\mu_{i}-\mu_{d}\right)  e^{\phi}+\mu_{d}%
-\mu_{i}n_{b}-n_{i},$ where $n_{\alpha},$ $v_{\alpha},$ denoting the number
density and speed of $\alpha-$species particle with $\alpha=i$ (for plasma
ions) and $b$ (for ion beams) are normalized by their equilibrium values
$n_{\alpha0}$ and the ion-sound speed $c_{s}\equiv\sqrt{k_{B}T_{e}/m_{b}}.$
Here $k_{B}$ is the Boltzmann's constant, \ $T_{e}$ is the electron
temperature and $m_{\alpha}$ is the mass of $\alpha-$species particle. Also,
$\phi$ is the electrostatic wave potential normalized by $k_{B}T_{e}/e,$ with
$e$ denoting the elementary charge, $\zeta_{(i,b)}=(\mu,1),$ where $\mu\equiv
m_{b}/m_{i}$ is the ion-beam to plasma ion mass ratio. Moreover, $\mu
_{d}=Z_{d}n_{d0}/n_{i0}$ is the ratio of the equilibrium dust density
(multiplied by $Z_{d},$ the number of electrons residing on the dust-grains)
to plasma ion density and $\mu_{i}=n_{b0}/n_{i0}$ is the ion-beam to plasma
ion density ratio. The space $(x)$ and time $(t)$ variables are respectively
normalized by the electron Debye length, $\lambda_{D}\equiv\sqrt{k_{B}%
T_{e}/4\pi n_{b0}e^{2}}$ and the inverse of the beam plasma frequency,
$\omega_{pb}\equiv\sqrt{4\pi n_{b0}e^{2}/m_{b}}$. At equilibrium, the overall
charge neutrality condition is $n_{e0}+Z_{d}n_{d0}=n_{i0}+n_{b0}.$

In order to derive the evolution equation for the propagation of small but
finite amplitude DIAWs, we use the standard reductive perturbation technique
(RPT) \cite{Washimi} in which the independent variables are stretched as
\ $\xi=\epsilon^{1/2}\left(  x-v_{p}t\right)  $ $\tau=\epsilon^{3/2}t$ . The
dependent variables, on the other hand, can be expanded as $n_{\alpha
}=1+\Sigma_{j=1}^{\infty}\epsilon^{j}n_{\alpha}^{(j)},$ $v_{\alpha}%
=v_{\alpha0}+\Sigma_{j=1}^{\infty}\epsilon^{j}v_{\alpha}^{(j)}$ with
$v_{i0}=0,$ and $\phi=\Sigma_{j=1}^{\infty}\epsilon^{j}\phi^{(j)},$ where
$\epsilon$ is a small nonzero constant measuring the weakness of the
dispersion and $v_{p}$ is the Mach number (phase speed of the DIAWs normalized
by the ion-sound speed, $c_{s}$).

Substituting the stretched coordinates and the expressions for $n_{\alpha},$
$v_{\alpha}$ and $\phi$ into the basic equations, and equating the
coefficients of different powers of $\epsilon$ we get from the lowest order of
$\epsilon$ the expressions: $n_{i}^{(1)}=\mu\phi^{(1)}/v_{p}^{2},$ $n_{b}^{(1)}=-\phi
^{(1)}/\left(  v_{p}-v_{b0}\right)  ^{2},$ $v_{i}^{(1)}=\mu\phi^{(1)}/v_{p},$
$v_{b}^{(1)}=-\phi^{(1)}/\left(  v_{p}-v_{b0}\right)  $ and $n_{i}^{(1)}%
+\mu_{i}n_{b}^{(1)}=\left(  1+\mu_{i}-\mu_{d}\right)  \phi^{(1)}\ $, together
with the dispersion law $v_{p}^{2}=\left(  \mu\delta^{2}-\mu_{i}\right)
/\left(  1+\mu_{i}-\mu_{d}\right)  \delta^{2},$ where $\delta=1-\gamma
\equiv1-v_{b0}/v_{p}$ is called the synchronism parameter such that
$v_{b0}=v_{p}$ for $\delta=0$. Proceeding in this way we finally obtain the
following Korteweg de Vries (KdV) equation%

\begin{equation}
\partial_{\tau}\phi+A\phi\partial_{\xi}\phi+B\partial_{\xi}^{3}\phi=0,
\label{a}%
\end{equation}
where $\phi\equiv\phi^{(1)},$ and the nonlinear coefficient $A$ and the
dispersive coefficient $B$ are given by $A=\left[  \left(  1+\mu_{i}-\mu
_{d}\right)  \delta^{4}v_{p}^{4}+3\left(  \mu_{i}-\mu^{2}\delta^{4}\right)
\right]  /2v_{p}\delta\left(  \mu_{i}-\mu\delta^{3}\right)  ,$ $B=-\delta
^{3}v_{p}^{3}/2\left(  \mu_{i}-\mu\delta^{3}\right)  .$ Note that for
$\delta=0$, $A$ becomes infinite and $B=0$. Also, for $\mu_{i}-\mu\delta
^{3}=0$, i.e., $\delta=(\mu_{i}/\mu)^{1/3}$, both $A$ and $B$ become infinite.
In this case, DIA soliton ceases to exist, since it is obtained as a solution
of Eq. (\ref{a}) where $A$ and $B$ are both finite and nonzero. The stationary
soliton solution of the KdV equation (\ref{a}) is obtained by transforming the
independent variables $\xi$ and $\tau$ to a single new variable $\zeta$
$=\xi-U_{0}\tau$, where $U_{0}$ is the constant phase speed (normalized by
$c_{s}$), and imposing the appropriate boundary conditions for localized
perturbations (\textit{viz}., $\phi\rightarrow0,\partial\phi/\partial
\xi\rightarrow0,$ $\partial^{2}\phi/\partial\xi^{2}\rightarrow0$ as
$\xi\rightarrow\pm\infty)$ as $\phi=\phi_{m}\sec h^{2}\left[  \left(
\xi-U_{0}\tau\right)  /D\right]  .$ The amplitude $\phi_{m}$ (normalized by
$k_{B}T_{e}/e$ ) and the width $D$ (normalized by $\lambda_{D}$) of the
soliton are given by $\phi_{m}=3U_{0}/A$ and $D=\sqrt{4B/U_{0}}.$

We note that $v_{p}$ is real either for $\gamma<1-\sqrt{\mu_{i}/\mu}$ (when
$0<\gamma<1$) or for $\gamma>1-\sqrt{\mu_{i}/\mu}$ (when $\gamma>1$). \ Also,
$B<0$ for $\gamma>1-\sqrt[3]{\mu_{i}/\mu}$, and then the width of the DIA
soliton becomes imaginary. So, for values of $\gamma$ in the regime
$1-\sqrt[3]{\mu_{i}/\mu}$ $<$ $\gamma$ $<1-\sqrt{\mu_{i}/\mu},$ the ion beams
will not be able to excite DIA solitons in our dusty ion-beam plasma.
Moreover, $B>0$ either for $\gamma<1-\sqrt[3]{\mu_{i}/\mu}$ or, for
$\gamma>1+\sqrt[3]{\mu_{i}/\mu}$, so that ion-beam driven DIA soliton
excitation is possible if and only if $\gamma$ satisfies either $\gamma$
$<$min$\{1-\sqrt{\mu_{i}/\mu},1-\sqrt[3]{\mu_{i}/\mu}\}$, i.e., $\gamma$
$<\gamma_{c1}\equiv1-\sqrt[3]{\mu_{i}/\mu}$ or, $\gamma>$max$\{1+\sqrt{\mu
_{i}/\mu},1+\sqrt[3]{\mu_{i}/\mu}\}$, i.e., $\gamma$ $>\gamma_{c2}%
\equiv1+\sqrt{\mu_{i}/\mu}$. Thus, there exist two critical values of $\gamma
$, namely $\gamma_{c1}$ and $\gamma_{c2}$ above and below which the ion-beam
driven DIA soliton does not exist. Both the critical values (one of which is
less than and other is always greater than the unity) depend on the ion-beam
to plasma ion density ratio as well as the ratio of their masses.
\begin{figure}[ptb]
\begin{center}
\includegraphics[height=1.5in,width=2.5in]
{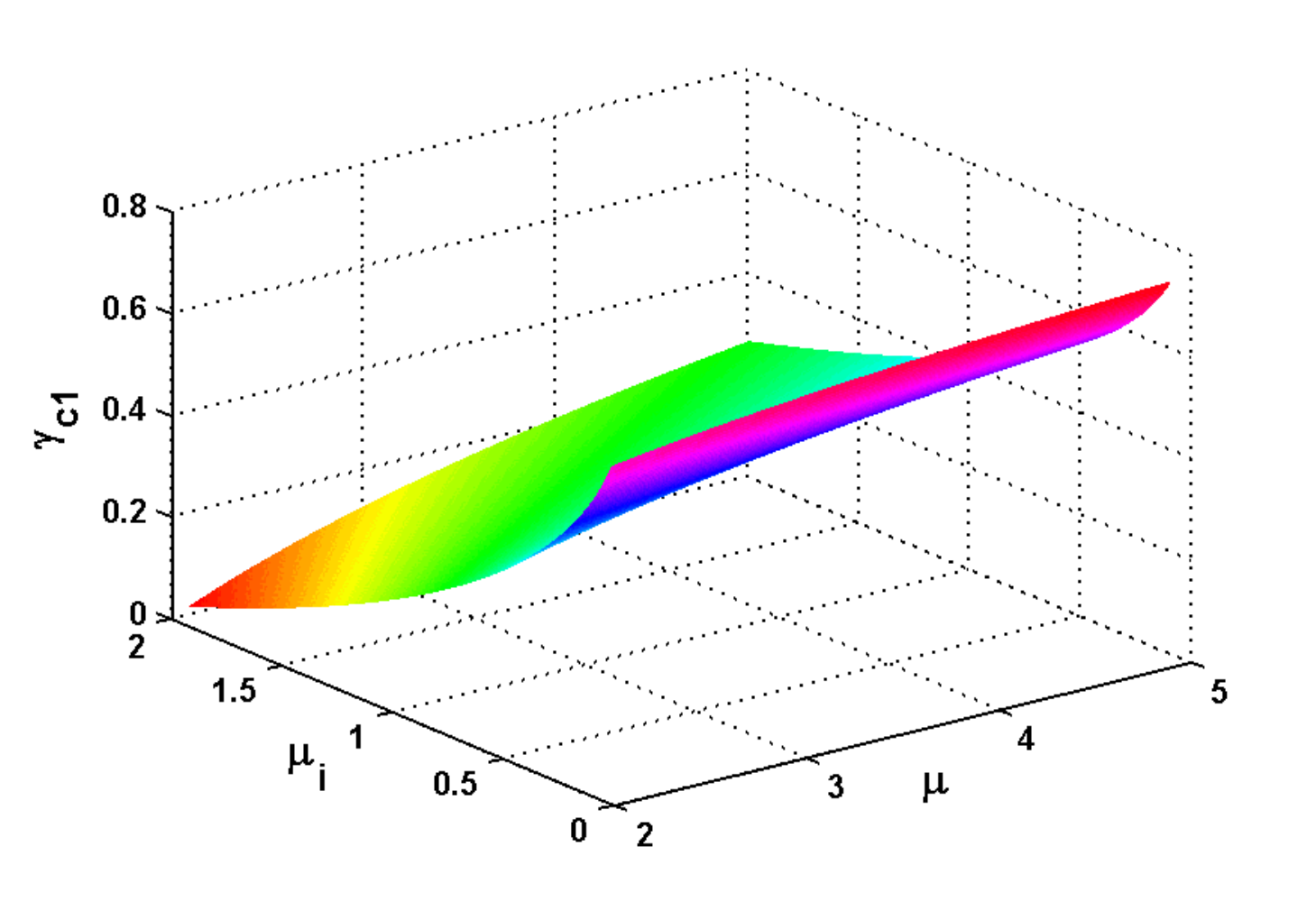}
\includegraphics[height=1.5in,width=2.5in]
{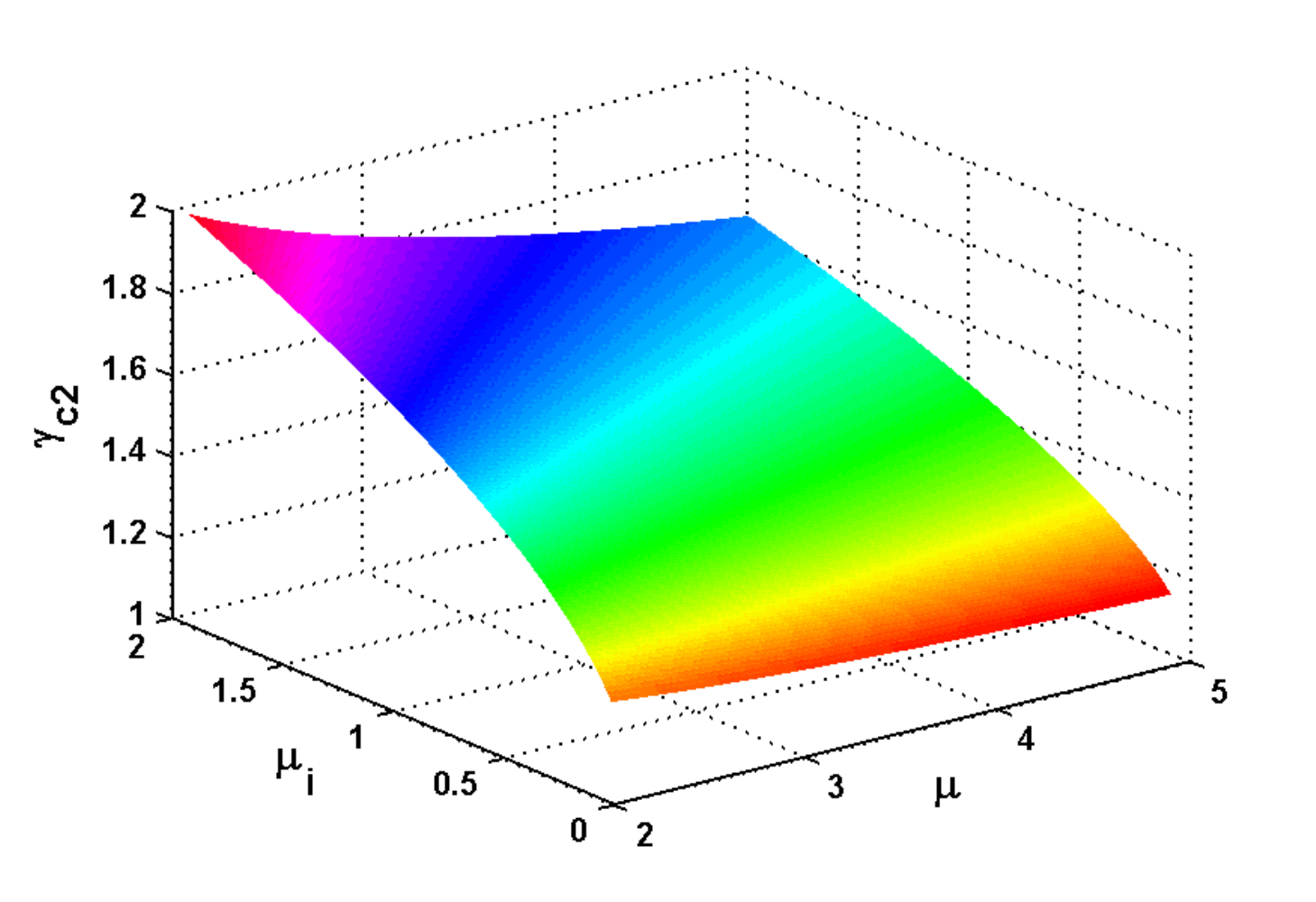}
\caption{(Color online) The critical values  $\gamma_{c1}$ (upper panel) and $\gamma_{c2}$ (lower panel) of $\gamma$ are shown  against $\mu$ and $\mu_i$.}
\end{center}
\end{figure}

We numerically investigate the properties of the critical values of $\gamma$,
the phase velocity $v_{p}$ as well as the nonlinear and dispersive
coefficients $A$ and $B$. Figure 1 shows that $\gamma_{c1}$ is always less
than unity and it increases with decreasing the values of $\mu_{i}$ and with
increasing $\mu$. Notice that $\mu>$ $\mu_{i}$, since $\gamma>0$. This means
that for the soliton solution to exist, if the ion-beam concentration
increases with respect to the plasma ion density, and the ion-beam mass
exceeds that of the plasma ions, the ion-beam speed has to be very small
compared to the phase speed. On the other hand, for relatively low-density of
ion-beams or the smaller values of the ratio $\mu_{i}$, the beam speed may
need to approach the phase speed for the excitation of solitons. In contrast
to the upper panel of Fig. 1, the lower panel shows that $\gamma_{c2}$ is
always greater than unity, and that the ion-beam speed must be larger than the
phase speed (when ion-beam density increases with respect to the plasma ion
concentration) in order to excite beam driven DIA solitons.
\begin{figure}[ptb]
\begin{center}
\includegraphics[height=1.5in,width=2.5in]
{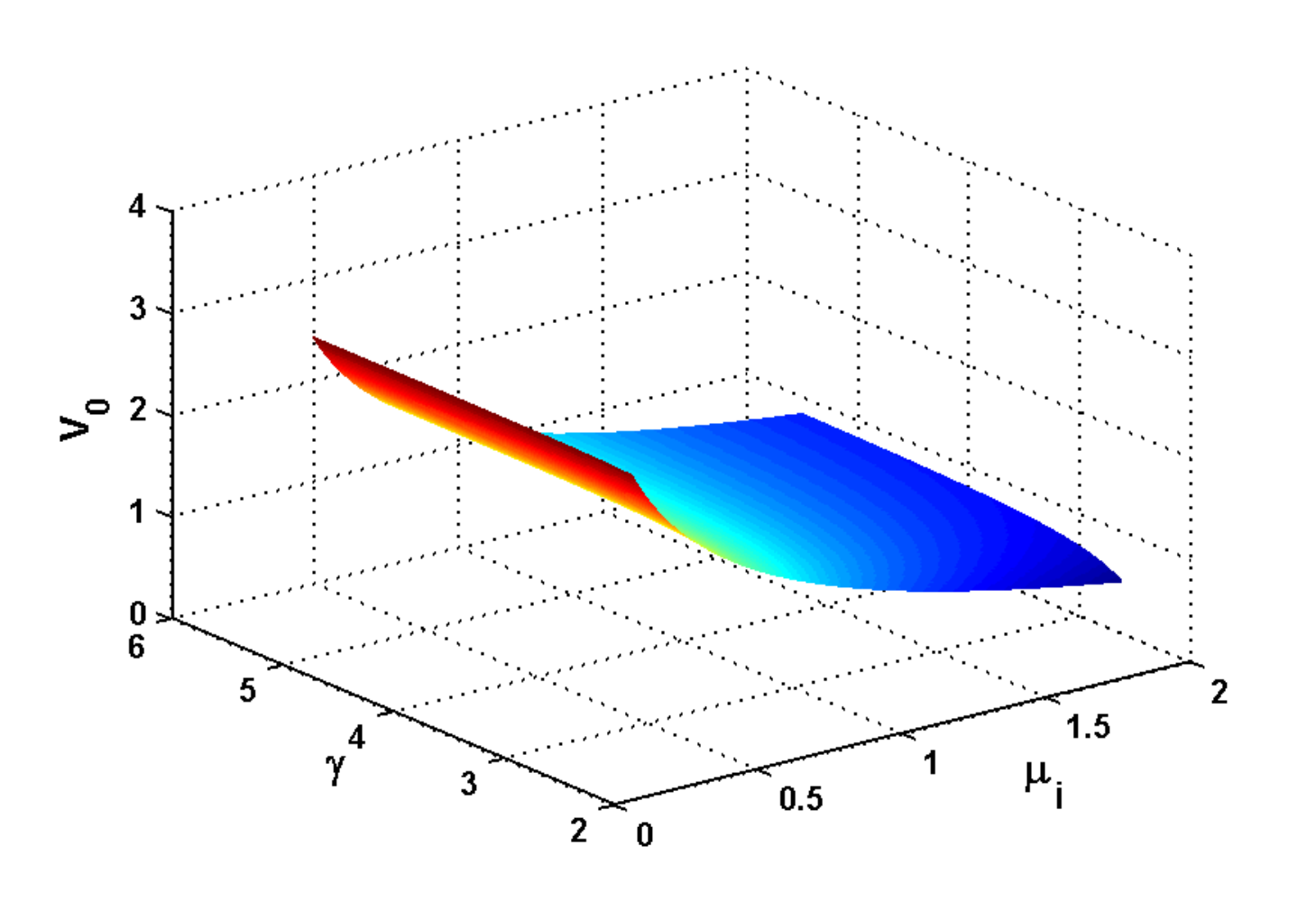}
\caption{(Color online) The phase speed $v_p$ is plotted against $\mu_i$ and $\gamma$ for $\mu_d=0.8, \mu=3$.}
\end{center}
\end{figure}

Figure 2 explains the behaviors of the phase speed $v_{p}$ with respect to
$\mu_{i}$, and $\gamma$ for $\mu=3, \mu_{d}=0.8.$ Since the ion-acoustic
solitary wave does not couple to that of ion beams, one can consider a range
of values of $\gamma$, which will enable to find $v_{p}$ quite easily. It is
seen that $v_{p}$ decreases with decreasing the ratio $\gamma$ and for
increasing the density ratio $\mu_{i}$. It also increases with increasing the
mass ratio $\mu$ as well as with increasing the impurity parameter $\mu_{d}$
(not shown in the figure), i.e. increasing the negative charge concentration
into the dust grains. Here we have considered an ion-beam moving in the
positive direction with a speed greater than the critical ion-beam speed,
i.e., $\gamma$ $>\gamma_{c2}$ (The case of $\gamma$ $<\gamma_{c1}$ is also
similar) for which $v_{p}$ and $B$ are real and finite. It is also found that
at low charged dust impurity, $v_{p}$ decreases faster the larger are the
ion-beam concentrations.
\begin{figure}[ptb]
\begin{center}
\includegraphics[height=1.5in,width=2.5in]
{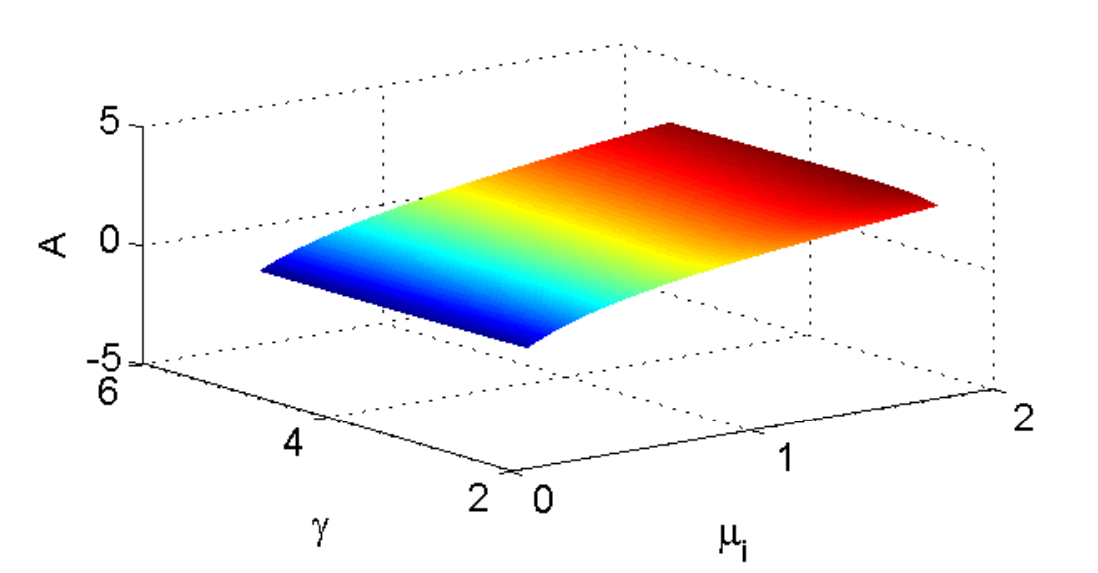}
\caption{(Color online) Dependence of the nonlinear coefficient $A$ on $\mu_i$ and $\gamma$ for $\mu_d=0.8, \mu=3$.}
\end{center}
\end{figure}
The variation of the nonlinear coefficient $A$ with respect to $\mu
_{i}$, and $\gamma$ for the same parameter values as in Fig. 2, is shown in
Fig. 3. We note that for a given value of $v_{p},$ the amplitude of the
beam-driven DIAWs depends on the various physical parameters, namely $\mu
_{i},\gamma,\mu$ and $\mu_{d}$. From Fig. 3, it is evident that the amplitude
of the DIA soliton is significant in the range $3<\gamma<4$. It decreases with
increasing the ion-beam concentration. Physically, as the number densities of
ion-beam increases compared to the plasma ions, the nonlinearity effect in the
system becomes higher and higher. On the other hand, the mass ratio $\mu$ is
also found to enhance the nonlinear coefficient $A$, and hence to decrease the
soliton amplitude. It is found that that as the negative charge concentration
on the dust grains decreases (or the nonlinear effects become larger) the
soliton amplitude also decreases. Moreover, we find that the effect of the
ratio $\gamma$ on the soliton amplitude is more pronounced at low charged dust concentration.
\begin{figure}[ptb]
\begin{center}
\includegraphics[height=1.5in,width=2.5in]
{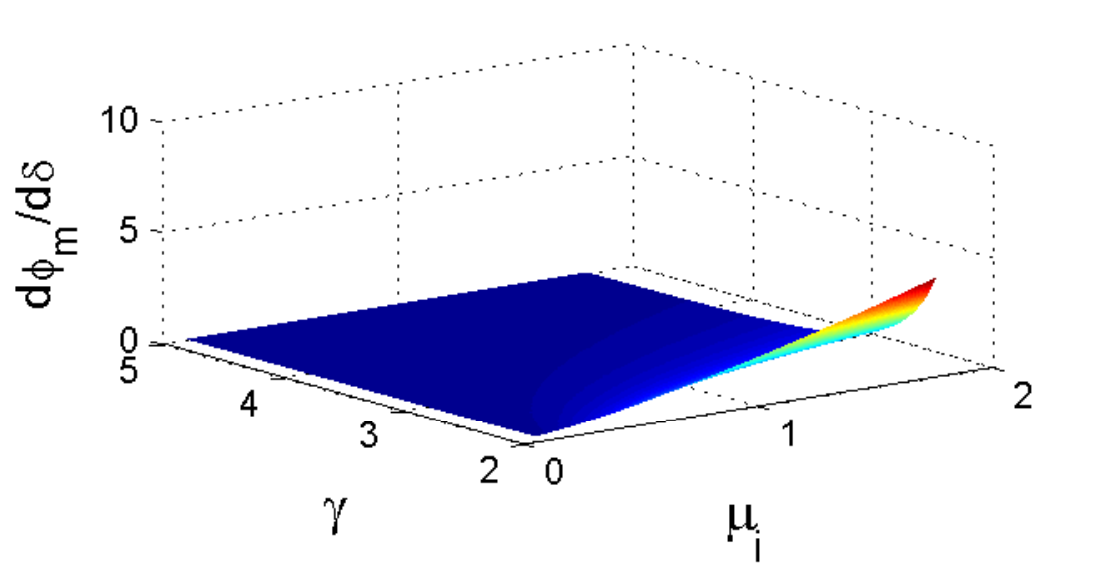}
\caption{(Color online) The absolute value of the beam amplification rate (where $\delta<0$) is shown with respect to $\mu$ and $\gamma$. The other parameters are as in Fig. 2.}
\end{center}
\end{figure}

Figure 4 shows that the absolute value of the beam amplification rate (since
$\delta=1-\gamma<0$ for $\gamma>1$) of soliton amplitude remains almost
unchanged as long as the ion-beam concentration remains less than the plasma
ion density, i.e., $\mu<1$. As the value of $\mu$ increases, the amplification
rate also increases, and it attains its maximum value at a higher ion-beam
concentration. The Latter turns out to the increase of the amplification rate
with decreasing values of $\gamma>\gamma_{c2}$. \ In the variations of the
dispersive coefficient $B$ with respect to the parameters as indicated above
we find that since, $B\propto D,$ the width of the soliton, for a prescribed
value of $v_{p}$ we can easily find the soliton widths with different plasma
parameters. As for example, Fig. 5 shows that for a fixed ion-beam mass and
constant charged dust concentration, the dispersive effects become stronger
with the beam speed. As a result, the width of the soliton increases with
increase of the ion-beam speed and reaches its maximum value. For relatively
higher values of the ion-beam concentration, the rate of decrease of the width
is comparatively high, while for \ lower values of the same, the ion-beam
density does not have much effect on the wave dispersion, and hence the
soliton width almost remains independent of it. It is found that at increasing
value of $\mu$ or for heavier ion-beams, the wave is more dispersive, and
hence the width becomes higher. By reducing the negative charge on the dust
grains, one can find much higher soliton width than that in Fig. 5. Also, at
higher ion-beam concentration, the width seems almost to be independent of
$\mu_{i}$. It is also evident that the rate of increase of the width with
respect to the ion-beam speed is\ more faster than the highly charged dust case.
\begin{figure}[ptb]
\begin{center}
\includegraphics[height=1.5in,width=2.5in]
{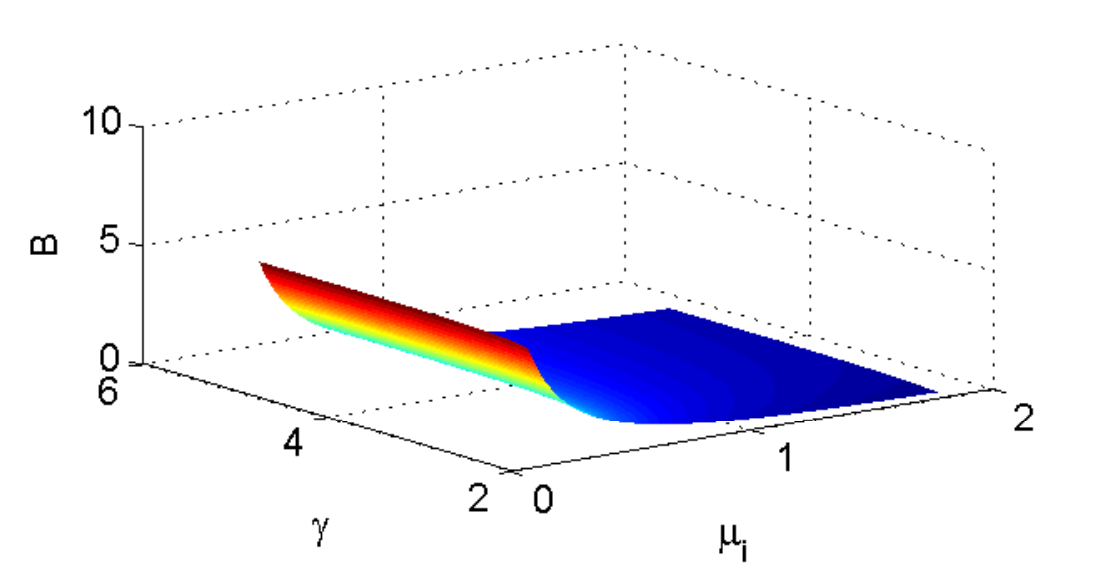}
\caption{(Color online) Dependence of the dispersive coefficient $(B)$ on $\mu_i$ and $\gamma$ for the parameter values as in Fig. 2.}
\end{center}
\end{figure}

Numerical solution of the KdV equation (\ref{a}) (see Fig. 6) shows that in
the very beginning (e.g., $\tau=0.5$), the sinusoidal positive pulse
propagates. As time progresses [e.g., $\tau=20,25$; see left and right panel
of Fig. 6], the leading part of the positive pulse gets steepened due to
nonlinearity and then as it travels more distance, the pulse breaks into a
train of solitons due to dispersion. The small hump in front of the peak [left
panel of Fig. 6] may be due to the reflected ions. The small hump appears
after the peak [right panel of Fig. 6] and it tends to disappear after a long
interval of time. Once the solitary peaks are generated, they propagate
keeping their shapes unchanged due to nice balance of the nonlinearity and
dispersion. By changing the system parameter values one can observe different
solitary peaks at different positions.
\begin{figure}[ptb]
\begin{center}
\includegraphics[height=1.3in,width=1.6in]
{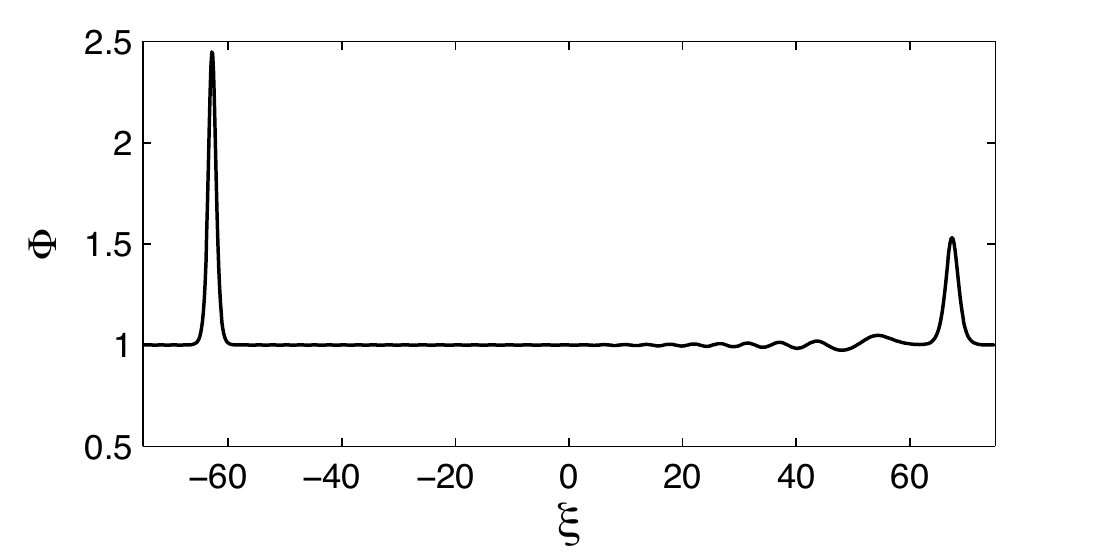}
\includegraphics[height=1.3in,width=1.6in]
{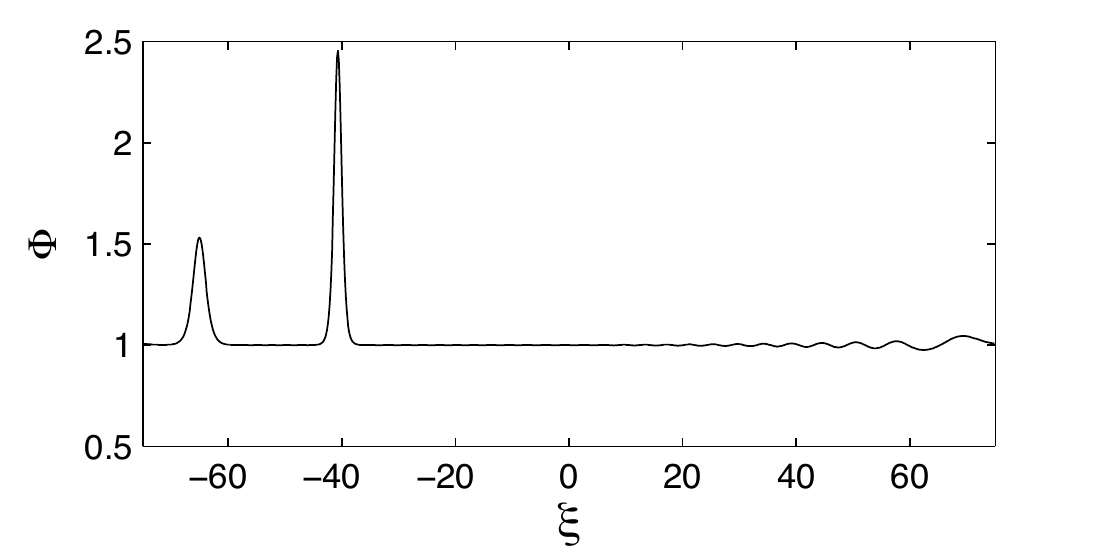}
\caption{Nonstationary soliton solution of the KdV equation with parameter values $\mu_i=1.6, \mu_d=0.8, \mu=\gamma=3$ at $\tau=20$ (left panel) and  $\tau=25$ (right panel), showing the space-time development of the DIA solitons for an initial Gaussian positive pulse of the form $\phi=1+\exp[-(\xi+2)^2/6)]$.}
\end{center}
\end{figure}

To summarize, we have investigated the nonlinear propagation of small but
finite amplitude DIA solitary waves in an ion-beam driven dusty plasma. The
conditions for the existence of such solitary waves as well as the properties
of the soliton amplitude and width in terms of the system parameters are
obtained and analyzed numerically. Two critical values of $\gamma$ have been
found beyond (above and below) which the formation of small amplitude DIA
solitons is not possible. The predicted results could be important for soliton
excitations in laboratory ion-beam driven plasmas as well as in space plasmas
(e.g., in the auroral regions) with or without immobile charged dust grains.
To conclude,  the DIA solitons in ion-beam driven plasmas with
stationary charged dust grains are quite distinctive from the usual
electron-ion plasmas, and may show experimentally the fascinating behaviors.
Works in this direction is underway, and will be communicated elsewhere.

\begin{acknowledgments}
A. P. M. gratefully acknowledges support from the Kempe
Foundations, Sweden.
\end{acknowledgments}


\begin{thebibliography}{99}                                                                                               %


\bibitem {Whipple}E. C. Whipple, T. G. Northrop and D. A. Mendis, J. Geophys.
Res. \textbf{90}, 7405 (1985).

\bibitem {Sheehan}D. P. Sheehan, M. Carilo, and W. Heidbrink, Rev. Sci.
Instrum. \textbf{61}, 3871 (1990).

\bibitem {Selwyn}G. S. Selwyn, J. Singh, R. S. Bennett, and J. Vac. Sci.
Technol. \textbf{A7}, 2758 (1989).

\bibitem {Matsumoto}H. Matsumoto, H. Kojima, T. Miyatake, Y. Omura, M. Okada,
I. Nagano, and M. Tsutui, Geophys. Res. Lett. \textbf{21}, 2915 (1994).

\bibitem {Shukla}P. K. Shukla and V. P. Silin, Phys. Scr. \textbf{45}, 508 (1992).

\bibitem {Barkan}A . Barkan, R. L. Merlino and N. D'Angelo, Phys. Plasmas
\textbf{2}, 3563 (1995).

\bibitem {Bharuthram}R. Bharuthram and P. K. Shukla, Planet. Space Sci.
\textbf{40}, 973 (1992); N. D'Angelo, Planet. Space Sci. \textbf{42}, 507
(1994); F. Sayed, M. M. Haider, A. A. Mamun, P. K. Shukla, B. Eliassson, and
N. C. Adhikary, Phys. Plasmas \textbf{15}, 063701 (2008).

\bibitem {Nakamura1}Y. Nakamura, H. Bailung, and P. K. Shukla., Phys. Rev.
Lett. \textbf{83}, 1602 (1999); N. C. Adhikary, M. K. Deka, and H. Bailung,
Phys. Plasmas \textbf{16}, 063701 (2009).

\bibitem {Okutsu}E. Okutsu, M. Nakamura, Y. Nakamura and T. Itoh, Plasma
Physics \textbf{20}, 561 (1978).

\bibitem {Temerin}M. Temerin, K. Cerny, W. Lotko, and F. S. Mozer, Phys. Rev.
Lett. \textbf{48}, 1175 (1982).

\bibitem {Parks}G. Parks, L. J. Chen, M. McCarthy, D. Larson, R. P. Lin, T.
Phan, H. Reme, and T. Sanderson, Geophys. Res. Lett. \textbf{25}, 3285, (1998).

\bibitem {Sugai}H. Sugai, T. H. Ahn, I. Ghanashev, M. Goto, M. Nagatsu, K.
Nakamura, K. Suzuki, and H. Toyoda, Plasma Phys. Controll. Fusion \textbf{39},
445 (1997).

\bibitem {Abrol}P. S. Abrol and S. G. Tagare, Phys. Lett. A. \textbf{75},14 (1979).

\bibitem {Nakamura2}Y. Nakamura and K. Komatsuda, J. Plasma Physics
\textbf{60}, 69 (1998); Y. Nakamura, H. Bailung and R. Ichiki, Phys. Plasmas
\textbf{11}, 3795 (2004).

\bibitem {Washimi}H. Washimi and T. Taniuti, Phys. Rev. Lett. \textbf{17}, 996
(1966).

\end{thebibliography}
\end{document}